\documentclass{eas}
\usepackage{graphicx}
\begin{document}

\title{Modelling the SEDs of spiral galaxies} 
\author{Cristina C. Popescu}\address{University of Central Lancashire, 
Preston, PR1 2HE, UK;
cpopescu@uclan.ac.uk}
\author{Richard J. Tuffs}\address{Max Planck Institut f\"ur Kernphysik,
  Saupfercheckweg 1, 69117 Heidelberg, Germany; Richard.Tuffs@mpi-hd.mpg.de}
\begin{abstract}
Modelling the UV/optical - infrared/submm SEDs of spiral galaxies
observed with Herschel will be an essential tool to quantitatively
interpret these observations in terms of the present and past
star-formation activity of these systems.
In this lecture we describe the SED modelling technique we have
developed, its applications and tests of its predictions. We show 
that both the panchromatic SED modelling of individual galaxies and the 
B-band attenuation-inclination relation of
large statistical samples suggest that spiral galaxies in the nearby Universe 
behave as optically thick systems in their global properties and large-scale
distribution of light (central face-on B-band opacity of ${\tau}^f_B\sim 4$). 
However disk galaxies are very
inhomogeneous systems, having both optically thick components (e.g. spiral
arms), and optically thin components (e.g. the interarm regions), the latter
making galaxies transparent to background galaxies.
\end{abstract}
\maketitle
\section{Introduction}

The measurement of star-formation rates (SFRs) and star-formation histories of
galaxies - and indeed of the universe as a whole - requires a quantitative
understanding of the effect of dust in attenuating the light from different
stellar populations and how the absorbed stellar light is re-emitted by the
dust in the infrared (IR). This can be achieved by modelling the whole spectral
energy distribution of galaxies (SED), from the UV/optical to the 
far-IR (FIR)/submm. Thus, a SED model is a tool to translate
observed SEDs of galaxies to their intrinsic properties, e.g. 
intrinsic distribution of stars and dust, opacity and SFRs. Such a tool will 
be crucial for the interpretation of data from the upcoming Herschel mission, 
which will continue the advancements in our knowledge of the FIR
properties of galaxies (see reviews of Tuffs \& Popescu 2004; Sauvage et al. 
2005 or Sauvage 2007). 

From a technical point of view an SED model needs to incorporate a radiative 
transfer calculation for a realistic distribution of stellar emissivity and 
dust opacity. A variety
of radiative transfer codes have been proposed in the literature: Kylafis \&
Bahcall (1987); Witt et al. (1992); Gordon et al. (2001); Bianchi et al. 
(1996); Baes \& Dejonghe (2001a,b); Baes et al. (2005a,b); Jonsson (2006); 
Bianchi (2007), just to mention those which have been developed mainly for 
application to galaxies (e.g. Byun et al. 1994; Witt \& Gordon 1996, 2000; 
Kuchinski et al. 1998, Ferrara et al. 1999, Pierini
et al. 2004). A comprehensive review of 
the different methods used in these
radiative transfer codes can be found in Kylafis \& Xilouris (2005). 
Another key element that any SED model needs to incorporate is a dust
model that would provide the optical properties of the dust grains and the
grain size distribution. The standard dust models that have been used were
those of Mathis, Rumple \& Nordsieck (1977), Draine \& Lee (1984), 
Dwek (1986), D{\'e}sert et al. (1990), Laor \& Draine (1993). More recently
these models have been revised and updated by Weingartner \& Draine (2001), 
Draine \& Li (2001), Li \& Draine (2001), Zubko et al. (2004), 
Draine \& Li (2007). Radiative transfer and/or  dust models can then
be employed to model the SEDs of galaxies in different wavelength ranges:
models which solely involve radiative transfer calculations have been used to account 
for the optical appearance of galaxies (Xilouris et al. 1997, 1998, 1999, 
Matthews \& Wood 2001, Bianchi et al. 2007), techniques which solely involve 
dust 
models have been used to account for the FIR/submm SEDs (Dale \& Helou 2002, 
Draine \& Li 2007), and techniques which calculate the transfer of radiation 
 in combination with a dust model have been used to
self-consistently account for both the UV/optical and FIR/submm SEDs (Silva et
al. 1998, Bianchi et al. 2000, Popescu et al. 2000). A review of the different
modelling techniques can be found in Popescu \& Tuffs (2005) for the case of
spiral galaxies, which is the subject of this lecture. A review of the
SED modelling of starburst galaxies can be found in Dopita
(2005) and of dwarf galaxies in Madden (2005)(see also Madden 2007, this
volume). In this lecture we illustrate the principles and techniques
of SED modelling by describing the particular model we have developed (Popescu
et al. 2000), together with selected applications.

\section{Description of the model}

Star-forming galaxies are fundamentally inhomogeneous,
containing highly obscured massive star-formation regions, as well as more 
extended structures harbouring older stellar populations which may 
be transparent or have intermediate optical depths to starlight.
Accordingly, our model divides the stellar population into an 
``old'' component (considered to dominate the output in B-band 
and longer wavelengths) and a ``young'' component 
(considered to dominate the output in the non-ionising UV).

The ``old'' stellar population can be constrained from resolved optical and 
near-IR images via the modelling procedure of Xilouris et al. (1999). 
The procedure uses the technique for
solving the radiation transfer equation for direct and multiply scattered
light for arbitrary geometries by Kylafis \& Bachall (1987). For edge-on 
systems these calculations completely determine the scale heights and lengths 
of exponential disk representations of the old stars (the ``old stellar disk'')
and associated diffuse dust (the ``old dust disk''), as well as a 
dustless stellar bulge. This is feasible for edge-on
systems since the scale height of the dust is less than that of the 
stars. These calculations also determine the central 
face-on opacity  (${\tau}^{f,1}_B$) of the  ``old dust disk''.
 The calculation is done independently for each optical/NIR image,
thus determining the extinction law for diffuse dust empirically.

The ``young'' stellar population is also specified by an exponential
disk, which we shall refer to as the ``young stellar disk''. 
Invisible in edge-on systems, 
its scale height is constrained to be $90\,$pc (the value for the Milky Way)
and its scale length is equated to that of the ``old stellar disk'' in B-band.
A second exponential dust disk  - the ``second dust disk'' is associated with 
the young stellar population. The dust associated
with the young stellar population was fixed to have the same scalelength and
scaleheight as for the young stellar disk. The reason for
this choice is that our thin disk of dust was introduced to mimic the diffuse
component of dust which pervades the spiral arms, and which occupies
approximately the same 
volume as that occupied by the young stars. This choice is also physically
plausible, since the star-formation rate is closely connected to the gas
surface density in the spiral arms, and this gas bears the grains which caused
the obscuration. 
Because two disks of dust are required 
for the model, we refer to it as the ``two-dust-disk'' model.
A schematic picture of these geometrical components of the model is given in
Fig.~1, together with a mathematical prescription of the stellar emissivities
and dust opacities used in the model.

\begin{figure}[htb]
\includegraphics[scale=0.5]{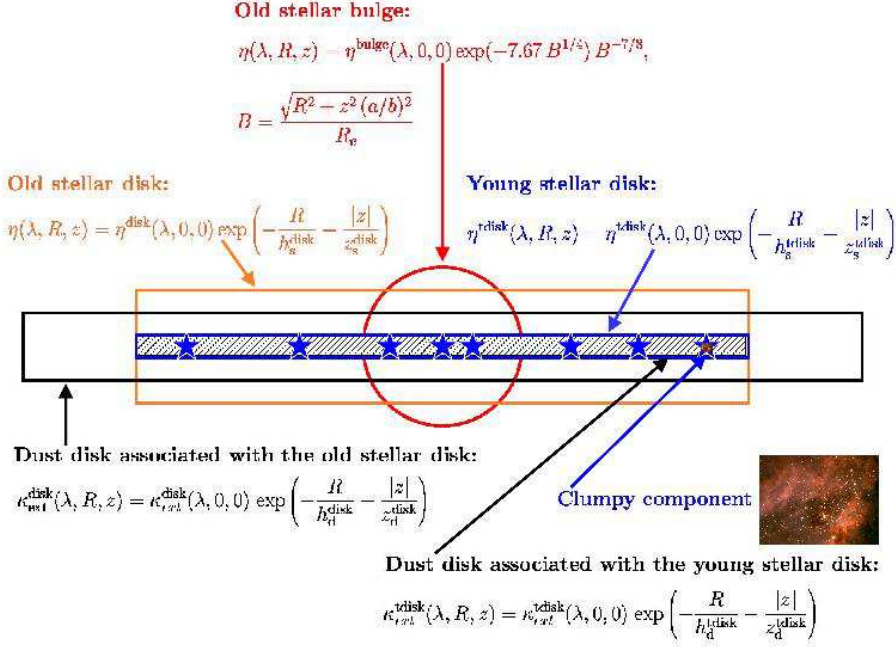}
\caption{Schematic representation of the geometrical distributions of stellar
  and dust emissivity used in the SED model of Popescu et al. (2000).} 
\end{figure}

The total UV output of the galaxy (expressed as the current star-formation 
rate ($SFR$)) and the central face-on opacity of the second
dust disk (${\tau}^{f,2}_B$)  are the first two primary 
free parameters of the model to determine the FIR/submm radiation.
They both relate to the smooth distribution of stars and dust
in the second disk. A third primary parameter, $F$, is included to
account for inhomogeneities in the distributions of dust and stars, in
particular due to star-forming regions.
By their very nature, star-forming regions harbour optically thick clouds which
are the birth places of massive stars. There is therefore a certain probability
that radiation from massive stars will be intercepted and absorbed by their 
parent clouds. This process is accounted for by the clumpiness factor $F$ 
which 
is defined as the total fraction of UV light which is locally absorbed in the
star-forming regions where the stars were born, giving rise to a local source
of warm dust emission. Astrophysically this process 
arises
because at any particular epoch some fraction of the massive stars have not had
time to escape the vicinity of their parent molecular clouds. Thus, $F$ is
related to the ratio between the distance a star travels in its lifetime due to
it's random velocity and the typical dimensions of star-forming complexes.
To conclude, in our formulation the clumpy distribution of dust is 
associated with the opaque parent molecular clouds of massive stars (see again 
Fig.~1 for a pictorial description of the diffuse and clumpy components of the 
model). In reality,
some of the dust in the diffuse dust disks may also be in clumps without
internal photon sources, but,
provided these clumps are optically thin, the transfer of radiation through
the disks will be virtually identical to that in a homogeneous disk. On
physical grounds, this condition will most often be met, since as soon as 
clumps become optically
thick to the impinging UV light, they will lose their principal source of heating
(the photoelectric effect) and will be prone to collapse and form stars
(Fishera \& Dopita 2007). This presumption is supported by the recent findings
of Holwerda et al. (2007a,b), that the structure of the diffuse ISM consists 
of optically thin dusty clouds.

The attenuation of the light from massive stars
by their parent molecular clouds has a different behaviour than the attenuation
of their light by the diffuse component. One difference is that the attenuation
by the parent molecular clouds is independent of the inclination of the
galaxy. Another 
difference is that the
wavelength dependence is not determined by the optical properties of the
grains (because the clouds are so opaque that they block the same proportion of
light from a given star at a given time at each wavelength), but instead 
arises because stars of different masses survive for
different times, such that lower mass and redder stars can escape further from
the star-forming complexes in their lifetimes. A proper treatment of 
the clumpiness factor is important since the clumpiness will change the
wavelength dependence of the UV attenuation curves of star-forming 
galaxies and will also affect the shape of the FIR SEDs.

After a further radiation transfer calculation (now
 incorporating the second dust disk) for the UV/optical/NIR light the heating 
of grains placed in the resulting radiation field 
can be determined. The illumination of the diffuse dust disks by 
optical/NIR photons is calculated by approximating the optical/NIR emissivity 
to that determined in the initial optical/NIR radiation transfer analysis, and 
is proportional to $SFR\,\times\,(1-F)$
for the non-ionising UV. The FIR-submm emission from 
grains for trial combinations of ${\tau}^{f,2}_B$, $SFR$, $F$ 
is then calculated for a grid of positions
in the galaxy. As described in detail in Popescu et al. (2000)
this calculation incorporates an explicit treatment of
the temperature fluctuations undergone by small grains whose cooling
timescales are shorter than the typical time interval between impacts
of UV photons. This so-called ``stochastic heating'' process determines
the amplitude and colour of the bulk of the diffuse emission from
most spiral galaxies in the shortest wavelength bands ($< 100\,\mu$m)
of Herschel. 
Subsequently we integrate over the entire galaxy to obtain the FIR-submm 
SED of the diffuse disk emission.
Prior to comparison with observed FIR-submm SEDs, an empirically
determined spectral template for the warm dust emission from the star-forming
 regions, scaled according 
to the value of $F$, must be added to this calculated spectral 
distribution of diffuse FIR emission.

Due to the constraints on the distribution of stellar emissivity in 
the optical-near infrared (NIR) and on the distribution and opacity
of dust in the ``old dust disk'' yielded by the radiation 
transfer analysis of the highly resolved optical-NIR images, coupled with the
simple assumptions for the distribution of the young stellar 
population and associated dust, our model has just three free 
parameters - $SFR$, $F$ and ${\tau}^{f,2}_B$. These fully determine
the FIR-submm SED, and allow a meaningful comparison with broad-band
observational data in the FIR/submm. As an example of a successful application
of the model we show the model fit to the SED of NGC891 (Fig.~2). Further
applications of the model can be found in Misiriotis et al. (2001). The models
have been also generalised to predict the attenuation of stellar light (Tuffs
et al. 2004) and the effect of dust on the observed scalelength and
central surface-brightness (M\"ollenhoff et al. 2006) of disk galaxies as a 
function of inclination, wavelength and total central face-on opacity 
${\tau}^f_B = {\tau}^{f,1}_B +{\tau}^{f,2}_B$.

\begin{figure}[htb]
\includegraphics[scale=0.6]{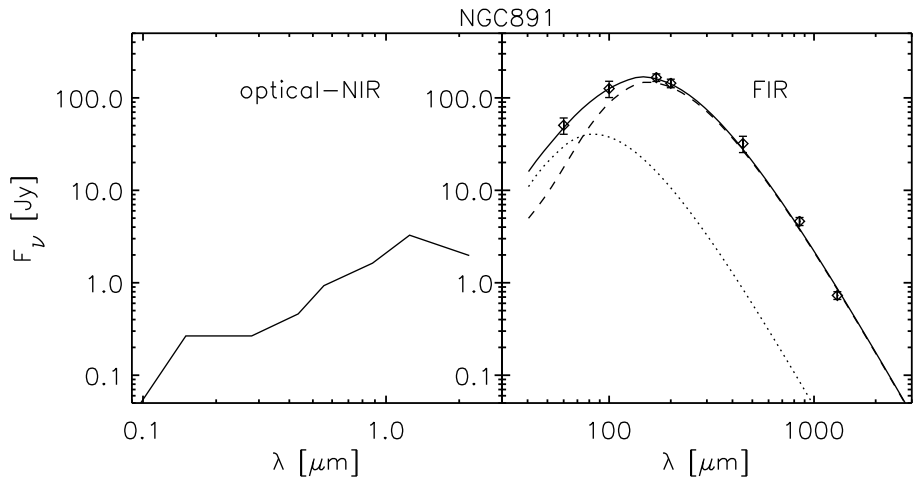}
\caption{The predicted SED of NGC~891 from the ``two-dust-disk'' model (Popescu
  et al. 2000).} 
\end{figure}

\section{The opacity derived from the SED modelling}

The central face-on opacity derived from the SED modelling has two terms. The
first term has a value ${\tau}^{f,1}_B \sim 1$, and characterises the 
optically thin
component of a spiral galaxy, which is the interarm region of the disk, or the
first disk of dust in our model. The second term has a value 
${\tau}^{f,2}_B \sim 3$, and represents the effective optical depth of the 
dust in the spiral arms, if this dust were distributed in a thin disk - the
second disk of dust in our model. Clearly the second disk of dust is not a real
element, meaning that the dust is not smoothly distributed in this disk. 
Dasyra et al. (2005) and Bianchi et al. (2007) showed, as expected, that the 
second disk of dust of our model cannot have a smooth distribution, otherwise 
it would have been seen as such in the near-infrared images of edge-on 
galaxies. The approximation would
obviously not work if one needs to predict the detailed arm-interarm contrast
of face-on systems.  However approximating the spiral distribution with a 
second disk of dust seems to be a very good approximation for predicting the 
major observational characteristics of spiral galaxies, as demonstrated by the
success of the model to pass several other observational tests.


\section{Testing the models}
\subsection{The surface brightness distribution in the FIR/submm}

We have seen that the model presented here can predict the optical appearance
of galaxies and the integrated FIR/submm SEDs. But a more stringent test is 
to see if it can also predict the appearance of galaxies in
the FIR/submm. In Fig.~3 one can see the comparison between the observed and
predicted radial profiles of the diffuse component of NGC891 at 170 and 
200\,${\mu}$m (Popescu et al. 2004). A very good agreement has been achieved 
between the observations and the model predictions. 
The small asymmetry in the observed profiles, at the level of about 
$10\%$ of the total emission, is consistent with our
model predictions for the emission from localised sources. 

\begin{figure}[htb]
\includegraphics[scale=0.55]{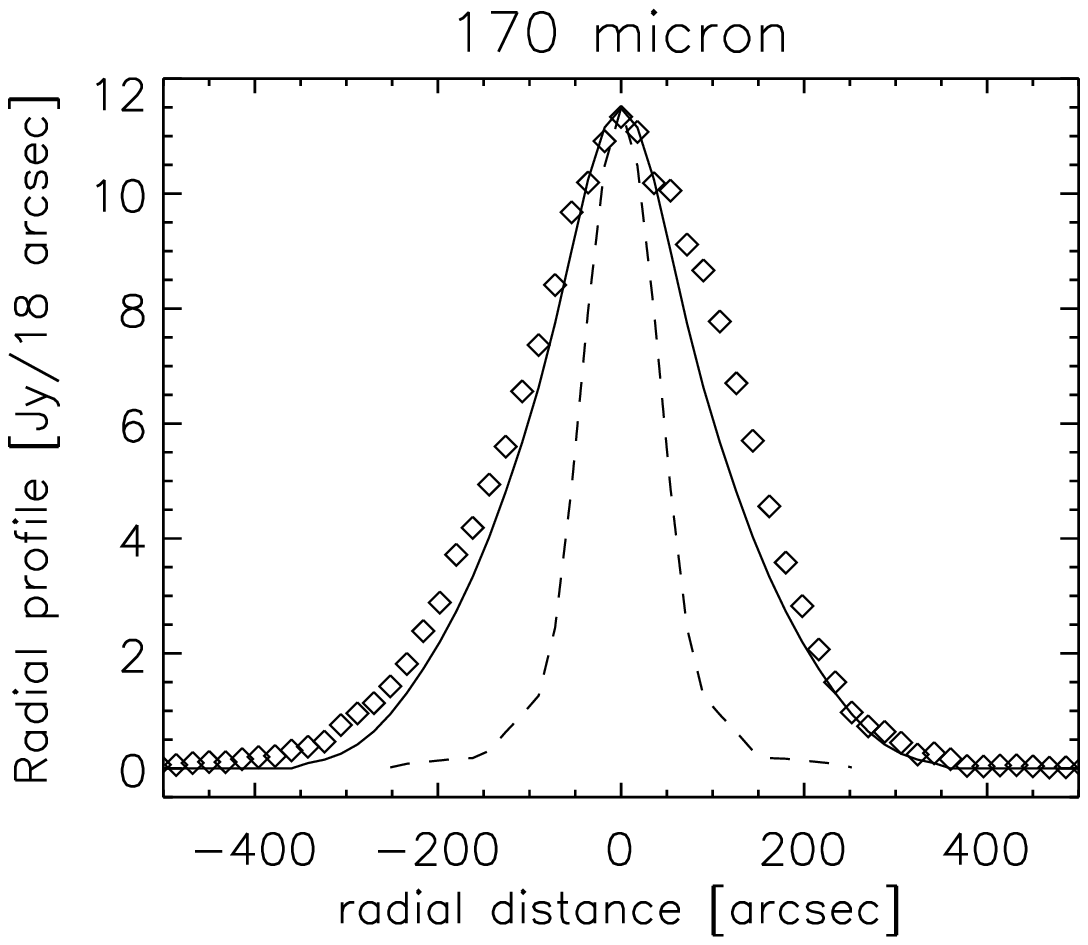}
\includegraphics[scale=0.55]{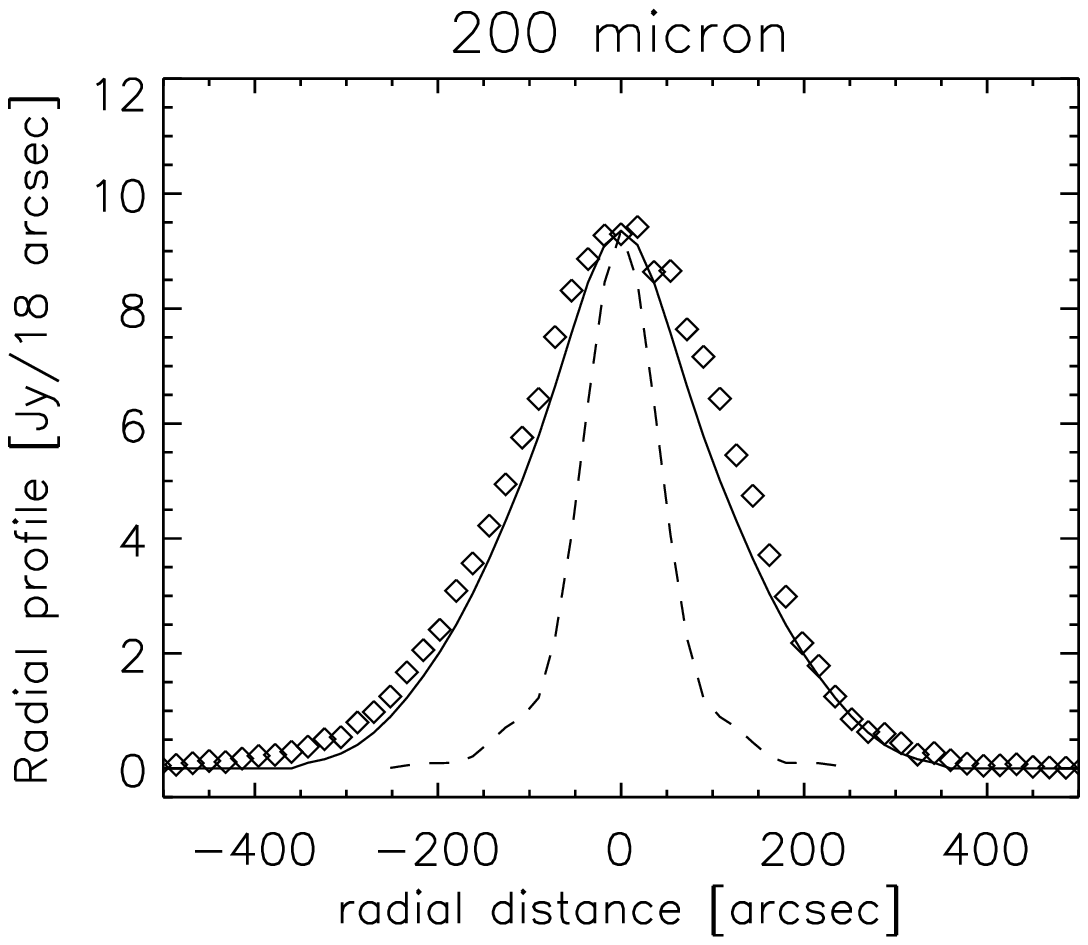}
\caption{Comparison between the observed (symbols) and predicted (solid line)
  FIR radial profiles of NGC891 (Popescu et al. 2004). The dotted lines
  represent the beam profile. The observations were done with the ISOPHOT
  instrument on board ISO. The predicted profiles are only for the diffuse
  component. The small excess emission in the observed profiles is likely to be
  due to localised sources.} 
\end{figure}

\subsection{The attenuation-inclination relation}

A fundamental test for our SED model is to see whether the typical 
solutions for the distribution of stars and dust and for the central face-on 
opacity needed to fit the panchromatic SEDs of individual galaxies can also 
predict the relation between optical attenuation and inclination derived
from observations of large statistical samples of galaxies.
This is an especially sensitive test, as
the rise in attenuation with inclination will very strongly depend on the
relative scaleheights of the assumed dust layers and stellar populations. In
particular it is an independent test for the existence of the second 
component of dust represented by the second dust disk. 
Historically, there has been some debate as to whether studies of galaxies at 
various inclinations can be used to constrain the dust distribution and 
opacity simultaneously (Holmberg 1958; Disney, Davies \& Phillipps 1989; 
Valentijn 1990; Disney et al. 1992). The general consensus was that one could 
not (see Davies \& Burstein 1995). Now we are in the position to re-address 
such an analysis, since the distribution of stars and dust has already been 
constrained in our model from the optical and FIR data, and the 
attenuation-inclination relation can be used as a consistency check for the 
solution found from the SED modelling.

\begin{figure}[htb]
\includegraphics[scale=0.35]{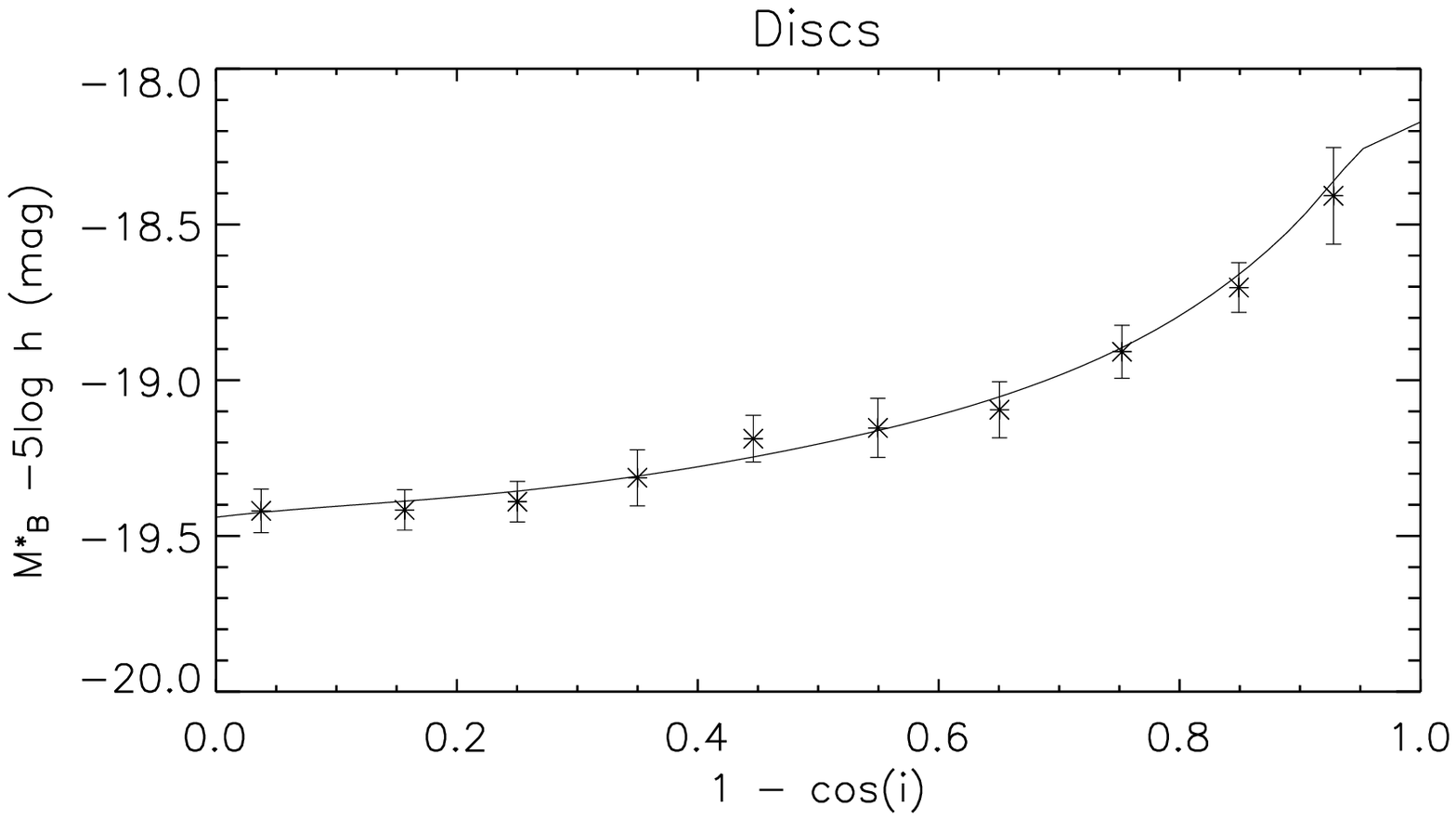}
 \includegraphics[scale=0.35]{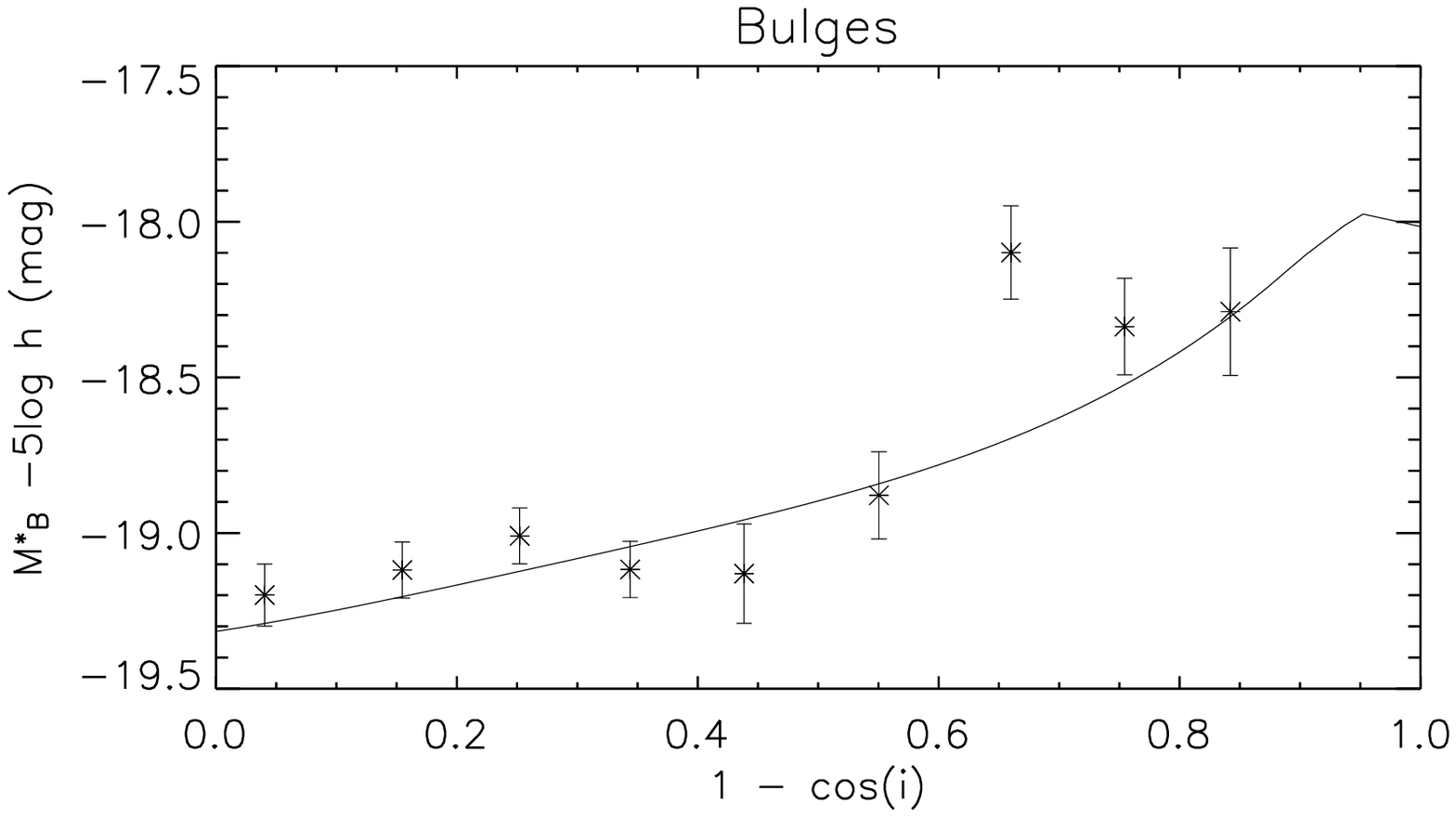}
\caption{The disk (left-hand panel) and bulge (right-hand panel) empirical
  attenuation-inclination relation derived from the MGC      
  Survey compared to the prediction of the dust model of Tuffs et al. (2004)
  with a central face-on B-band opacity ${\tau}^{f}_{B} = 3.8$, as derived
  from the best-fitting solution for the attenuation-inclination relation for
  disks (Driver et al. 2007).} 
\end{figure}

To this end we compared our model predictions with the B-band
attenuation-inclination relation derived from the Millennium
Galaxy Catalogue (MGC; Liske et al. 2003). The MGC is ideal for such
an analysis due to its unique depth and spectroscopic completeness,
and because all MGC galaxies have bulge-disk decompositions (Allen et al.
2006), which allowed the derivation of the attenuation-inclination
relations separately for disks and bulges. These relations are
shown in Fig.~4 together with the prediction from our SED model. 
One can see that the model predicts the data
rather well. The best-fitting solution for the disks is for a
central face-on opacity of $\tau^{f}_{B} = 3.8 \pm 0.7$, a solution that can
simultaneously account for the attenuation-inclination relation of the bulge as
well and is consistent with the solution found from the SED modelling of
individual galaxies. Fuller details of this analysis are given in Driver et 
al. (2007).

\begin{figure}[htb]
\includegraphics[scale=0.35]{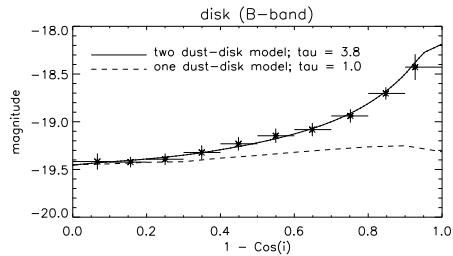}
\caption{Comparison between the predictions of the two dust-disk model (Popescu
  et al. 2000) with ${\tau}^{f}_{B}=3.8$ (solid line) and of the one-dust disk 
model (Xilouris et al. 1997, Alton et al. 2004, Dasyra et al. 2005, 
Bianchi et al. 2007), with
  ${\tau}^{f}_{B}=1$ (dashed line) for the attenuation-inclination relation for
  disks. The empirical attenuation-inclination relation derived from the MGC 
  is plotted as symbols. One can see that the one-dust disk model fails to
  reproduced the observed data.}
 
\end{figure}

We should mention here that one of the strength of this test is that it is
completely independent of the assumed dust emission properties. Thus the 
success of the solution found from the FIR/submm in predicting the
attenuation-inclination relation should also remove the degeneracy between
opacity and enhanced submm emissivity of the dust in the diffuse dust disks. 
In particular Alton et al. (2004), and 
Dasyra et al. (2005) have suggested that a solution with only one disk of dust 
and ${\tau}^{f}_{B}=1$, as derived from fitting the optical images of edge-on
galaxies (Xilouris et al. 1997, 1998, 1999, Bianchi et al. 2007), would still
reproduce the FIR SEDs of galaxies if the dust emissivity were enhanced over
what is assumed in typical dust models (e.g. Draine \& Lee 1984).
In Fig.~5 we plot the prediction for the attenuation-inclination relation
of the model with one disk of dust and ${\tau}^{f}_{B}=1$. One can see that
this model fails to predict the steep rise in attenuation shown by the data.
This indicates that it is this second dust-disk, which mimics the dust in the
spiral arms, that is needed to reproduce the FIR/submm SEDs of galaxies and the
steep rise in attenuation with inclination.

\section{Are spiral galaxies optically thin or optically thick?}

The value of ${\tau}^{f}_{B}\sim 4$ obtained from both the SED modelling of
individual galaxies and the attenuation-inclination relation of large
statistical samples suggest that spiral galaxies behave as optically thick
systems in their global properties. This is however fully consistent with the
spiral disks having their interarm regions transparent to background
galaxies (White et al. 2000). In other words the spiral disks are overall 
optically thick systems, but have an interarm component which is optically 
thin when viewed face-on. 



\begin{thebibliography}{99}
\bibitem[Allen et al. 2006]{mgc08} Allen P., Driver S.P., Graham A.W., Cameron
  E., Liske J. et al. 2006, MNRAS 371, 2
\bibitem[]{} Alton, P.B., Xilouris, E.M., Misiriotis, A., Dasyra, K.M., \&
  Dumke, M. 2004, A\&A 425, 109 
\bibitem[]{} Baes, M., \& Dejonghe, H. 2001a, MNRAS.326, 722
\bibitem[]{} Baes, M., \& Dejonghe, H. 2001b, MNRAS.326, 733
\bibitem[]{} Baes, M., Dejonghe, H., \& Davies, J.I. 2005a, Eds. C.C. Popescu 
and R.J. Tuffs, AIPC 761, 27
\bibitem[]{} Baes, M., Stamatellos, D., Davies, J.I., Whitworth, A.P., 
Sabatini, S. et al. 2005b, NewA 10, 523
\bibitem[]{} Bianchi, S. 2007, A\&A in press (arXiv:0705.1471v1)
\bibitem[]{} Bianchi, S., Ferrara, A., \& Giovanardi, C. 1996, ApJ 465, 127
\bibitem[]{} Bianchi, S., Davies, J.I. \& Alton, P.B. 2000, A\&A 359, 65
\bibitem{} Byun, Y. I., Freeman, K. C. \& Kylafis, N. D. 1994, ApJ 432, 114
\bibitem[]{} Dasyra, K.M., Xilouris, E.M., Misiriotis, A., \& Kylafis,
  N.D. 2005, A\&A 437, 447
\bibitem[Davies \& Burstein 1995]{davies} Davies J.I., Burstein D., 1995, in {\it The opacity of spiral discs}, NATA ASI, (Publ: Kluwer)
\bibitem[]{} D\'esert F.-X, Boulanger F., Puget J.L., 1990, A\&A 237, 215
\bibitem[Disney et al. 1989]{ddp90} Disney M.J., Davies J.I., Phillipps S.,
  1989, MNRAS 239, 939 
\bibitem[]{} Disney, M., Burstein, D., Haynes, M.P., \& Faber, S.M. 1992,
  Nature 356, 114
\bibitem[]{} Dopita, M.A. 2005, Eds. C.C. Popescu 
and R.J. Tuffs, AIPC 761, 203
\bibitem[]{} Draine B.T. \& Lee H.M. 1984, ApJ 285, 89
\bibitem[]{} Draine, B.T. \& Li, A. 2001, ApJ 551, 807
\bibitem[]{} Draine, B. T.\& Li, A. 2007, ApJ 657, 810
\bibitem[]{} Driver, S.P., Popescu, C.C., Tuffs, R.J., Liske, J., Graham, A.W.,
  et al. 2007, MNRAS 379, 1022
\bibitem{} Dwek E., 1986, ApJ 302, 363
\bibitem{} Ferrara, A., Bianchi, S., Cimatti, A., \& Giovanardi, C. 1999, 
ApJS 123, 437
\bibitem[]{} Fishera, J. \& Dopita, M.A 2007, ApJ submitted
\bibitem[]{} Gordon, K.D., Misselt, K.A., Witt, A.N., \& Clayton, G.C. 2001, 
ApJ 551, 269
\bibitem[Holmberg 1958]{holmberg58} Holmberg E., 1958, Medd. Lunds Obs. II,
  No. 136 
\bibitem[]{} Holwerda, B.W, Draine, B., Gordon, K.D., Gonzalez, R.A., Calzetti,
  D. et al. 2007a, AJ in press (arXiv0707.4165)
\bibitem[]{} Holwerda, B.W., Meyer, M., Regan, M., Calzetti, D., Gordon,
  K. D. et al. 2007b, AJ 134, 1655
\bibitem[]{} Jonsson, P. 2006, MNRAS 372, 2
\bibitem[]{} Kylafis, N. D. \& Bahcall, J. N., 1987, ApJ, 317, 637
\bibitem[]{} Kylafis, N.D. \& Xilouris, E.M. 2005, eds. C.C. Popescu \& R.J. 
Tuffs, AIPC 761, 3 
\bibitem{} Kuchinski, L. E., Terndrup, D. M., Gordon, K. D., \& Witt,
A. N. 1998, AJ 115, 1438
\bibitem{} Laor A., Draine B.T., 1993, ApJ 402, 441
\bibitem[]{} Li, A. \& Draine, B.T. 2001, ApJ 554, 778
\bibitem[1994]{alref1} Liske, J., Lemon, D., Driver, S.P., Cross, N.J.G., \&
  Couch, W.J.  2003, MNRAS, 344, 307
\bibitem[]{} Madden, S.C. 2005, Eds. C.C. Popescu and R.J. Tuffs, AIPC 761, 223
\bibitem[]{} Madden, S.C. 2007, this volume
\bibitem{} Mathis J.S., Rumpl W., Nordsieck K.H. 1977, ApJ 217, 425
\bibitem[]{} Misiriotis A., Popescu, C.C., Tuffs, R.J., \& Kylafis, N.D. 2001, 
A\&A, 372, 775
\bibitem[]{} M\"ollenhoff, C., Popescu, C.C. \& Tuffs, R.J. 2006, A\&A 456, 941
\bibitem[]{}
Pierini, D., Gordon, K.D., Witt, A.N., Madsen, G.J. 2004, ApJ 617, 1022
\bibitem[]{} Popescu, C.C. \& Tuffs, R.J. 2005, Eds. C.C. Popescu and
  R.J. Tuffs, AIPC 761, 155
\bibitem[]{} Popescu, C. C., Misiriotis, A., Kylafis, N. D., 
Tuffs, R. J. \& Fischera, J., 2000, A\&A, 362, 138
\bibitem[]{pop04} 
Popescu, C.C., Tuffs, R.J., Kylafis, N.D. \& Madore, B.F. 2004,
A\&A 414, 45
\bibitem[]{} Sauvage, M. 2007, this volume
\bibitem[]{} Sauvage, M., Tuffs, R.J. \& Popescu, C.C. 2005, in 
``ISO science legacy - a compact review of ISO major achievements'', 
Space Science Reviews, eds. C. Cesarsky and A. Salama, 
Springer Science + Business Media, Inc., vol. 119, Issue 1-4, p. 313 
\bibitem[]{} Strauss, M.A. Weinberg, D.H., Lupton, R.H. et al. 
2002, AJ 124, 1810
\bibitem[]{} Tuffs, R.J. \& Popescu, C.C 2005, Eds. C.C. Popescu and
  R.J. Tuffs, AIPC 761, 344
\bibitem[]{}
Tuffs, R.J., Popescu, C.C., V\"olk, H.J., Kylafis, N.D., \& Dopita, M.A. 
2004, A\&A 419, 821
\bibitem[Valentijn 1990]{valentijn} Valentijn E.A., 1990, Nature, 346, 153
\bibitem[]{} Weingartner, J.C. \& Draine, B.T. 2001, ApJ 548, 296
\bibitem[]{} White III, R.E., Keel, W.C. \& Conselice, C.J. 2000, ApJ 542, 761
\bibitem[]{} Witt, A.N., Thronson, H.A. \& Capuano, J.M. 1992, ApJ 393, 611
\bibitem{} Witt, A. N., \& Gordon, K. D. 1996, ApJ 463, 681
\bibitem{} Witt, A. N., \& Gordon, K. D. 2000, ApJ 528, 799
\bibitem[]{} Xilouris, E.M., Kylafis, N.D., Papamastorakis, J., Paleologou 
E.V. \& Haerendel, G. 1997, A\&A, 325, 135
\bibitem[]{} Xilouris, E.M., Alton, P.B., Davies, J.I., et al., 1998, A\&A, 
331, 894
\bibitem[]{} Xilouris, E. M., Byun, Y. I., Kylafis, N. D., 
Paleologou, E. V., Papamastorakis, J., 1999, A\&A, 344, 868
\bibitem[]{} Zubko, V., Dwek, E. \& Arendt, R.G. 2004, ApJS 152, 211
\end{thebibliography}
\end{document}